\documentclass[12pt,a4paper]{article}
\usepackage{amsmath}
\usepackage{graphicx}
\usepackage{times}
\begin{document}
\begin{titlepage}
\title{Does diffraction cone shrinkage with energy originate from unitarity?}
\author{ S.M. Troshin, N.E. Tyurin\\[1ex]
\small  \it NRC ``Kurchatov Institute''--IHEP\\
\small  \it Protvino, 142281, Russian Federation}
\normalsize
\date{}
\maketitle

\begin{abstract}
We note that the diffraction cone shrinkage might result  from unitarization only, i.e. there is no need to introduce it into an input amplitude ab initio. 
\end{abstract}
\end{titlepage}
\setcounter{page}{2}
\section*{Introduction}
Application of the unitarization procedure is an economy and convenient way to construct a true scattering amplitude obeying unitarity, in particular, when an  input  amplitude is  not constrained by the unitarity. There are several different unitarization mechanisms generating the required final output (cf. for systematic comparison\cite{glushko} and references therein).  

The need for unitarization has become evident at the time when the total cross--sections rise has been discovered. To fit the Regge model to the experimental data one should introduce a Pomeron pole  contribution   having intercept $\alpha(0)$ greater than unity. Such contribution would finally violate unitarity and therefore requires unitarization. The input amplitude of the Regge model with linear trajectory $\sim (s/s_0)^{\alpha(t)}$, however, includes diffraction cone shrinkage ab initio, i.e. the slope parameter $B(s)$ increases with energy logarithmically, $B(s)\sim \alpha'(0)\ln (s/s_0)$, $\alpha'(0)\neq 0$, while unitarity requires its double logarithmic asymptotic growth, $B(s)\sim \ln^2(s/s_0)$ if the total cross-section saturate Froissart-Martin bound. To reconcile the speed of shrinkage with unitarity requirements and with asymptotic saturation of this bound, one can consider  the slope of the Pomeron trajectory $\alpha'(0)$ being an energy--dependent effective function in the course of phenomenological consideration (cf. \cite{rysk}). 

An increase of the slope $B(s)$ and its speeding up with energy are then due to both  the unitarization procedure  and the Regge parameterization  for the input amplitude; since the input amplitude itself implies growth of $B(s)$ with energy, it is difficult, therefore, to deal with the two sources of
the energy dependence.
The importance of such disentanglement for the studies of the hadron interaction dynamics and diffraction processes in soft region is evident.

Both the $s$- and $t$- dependencies of the slope of the diffraction cone $B(s,t)$ are under active discussion  nowadays in connection with the new elastic scattering measurements performed at the LHC \cite{totem}, which have posed new issues  addressed in a number of papers, e.g. \cite{anal,epja,jenk}.

This note is devoted to discussion of an alternative interpretation of the origin of $B(s)$ growth when it is a unitarity effect alone. No doubt, this  is a model-dependent result and it correlates with a  form  for the input amplitude used at the unitarization.  But, the set of such models is rather wide and includes all the models assuming a factorized $s$- and $t$- dependence of the input amplitude. It includes geometrical models operating with the amplitudes in the impact parameter representation
(cf. for definition \cite{heny}). In these models an input amplitude  is taken as an overlap of matter distributions $D_1  \otimes D_2$ of the colliding hadrons following to the pioneering paper by Chou and Yang \cite{chy}.  It should be noted that the above said  factorization  results also from  the tower diagrams calculations in electrodynamics \cite{cheng}.

 \section{Unitarization of a factorized input }
In our qualitative consideration   we  suppose  that the real part of the elastic scattering amplitude is vanishingly small and can be neglected\footnote{However, it should be noted that such assumption is not quite correct in view of dispersion relations and it can be corrected by the restoration of the real part of the scattering amplitude with the scenario described in \cite{anal}.} since the high energy experimental data are in favor of the pure 
imaginary amplitude. 
We discuss the slope of the diffraction cone
\begin{equation}\label{bs0}
B(s)=\frac{d}{dt}\ln \frac{d\sigma}{dt}|_{t=0},
\end{equation}
It  is determined by the mean value of the impact parameter squared $b^2$, 
\begin{equation}\label{aver}
\langle b^2\rangle=\frac{\int_0^\infty b^3dbf(s,b)}{\int_0^\infty bdbf(s,b)}.
\end{equation}

The studies of  geometrical properties of hadron interactions   are important  \cite{blg} for understanding of hadron dynamics ultimately related to the  development of QCD in its nonperturbative sector. 

Unitarity  allows variation of the impact parameter dependent amplitude $f(s,b)$ in the region
$0\leq f \leq 1$, while the assumption on the absorptive scattering mode reduces the region of elastic scattering amplitude  variation to the interval $0\leq f \leq 1/2$. The value of $f=1/2$ corresponding to the complete absorption of the initial state means that the elastic scattering matrix element  is zero, $S=0$  ($S=1-2f$).  Unitarization schemes can provide an output amplitude $f$ limited by the unitarity itself ($U$--matrix) or by the black disc limiting value of $1/2$ (eikonal, method of continued unitarity) \cite{glushko}. Mechanisms of generation diffraction cone slope increase with energy are similar for these unitarization schemes. 

Due to the above similarity, we consider particular one of the above schemes, namely $U$--matrix \cite{uma}.  In the $U$--matrix approach (in the pure imaginary case) the relation between the scattering amplitude and the input quantity $u$ is simple:
\begin{equation}\label{um}
f(s,b)=u(s,b)/[1+u(s,b)],
\end{equation}
where $u$ is non-negative. 

The geometrical models asssume that $u(s,b)$ has a factorized form
\begin{equation}\label{usb}
u(s,b)=g(s)\omega(b),
\end{equation}
where $g(s)\sim s^\lambda$, the power dependence guarantees asymptotic growth of the total cross--section $\sigma_{tot}\sim \ln^2 s$. This factorized form and Eq. (\ref{um})  and the particular form of the function $\omega(b)$ chosen in accordance with the analytical properties of the scattering amplitude lead also to the behavior
\begin{equation}\label{bs}
B(s)\sim \ln^2 s.
\end{equation}
As it was noted in the Introduction, a simple way to construct the function $\omega(b)$ is to represent it as a convolution of the matter distributions in transverse plane as it was proposed by Chou and Yang \cite{chy}:
\begin{equation}
\omega (b)\sim D_1\otimes D_2\equiv \int D_1({\bf b}_1)D_2({\bf b}-{\bf b}_1).
\end{equation}
This function can also be constructed taking into account the hadron quark structure \cite{chiral}. 
The form of the function $\omega (b)$ consistent with analyticity is linear exponent at large values of $b$ and the following form can be adopted for simplicity
\begin{equation}\label{omb}
\omega (b)\sim \exp{(-\mu b)}.
\end{equation}
The energy independent parameter $\mu$ is related to a particular chosen physics model and the hadron structure, it can be  assumed that   $\mu=2 m_\pi$.
A general issue is that the diffraction cone slope $B^0$ corresponding to this factorized input amplitudes does not depend on the collision energy. It is determined by the geometrical radii of colliding particles.  The geometrical radius of particle is determined by the minimal mass of the quanta whose exchange is responsible for the scattering \cite{yuk}. The energy dependence of final slope $B(s)$ is generated by the  unitarization itself. 
Simple physical interpretation based on the analogy with bremsstrahlung can be found in \cite{solo}.

It should be noted that the energy dependence of $B(s)$ appears  at any energy value, 
and it $\sim \ln^2 s$ at asymptotics. At moderately small energies, where $g(s)$ is small, it becomes  (cf. Eqs. (\ref{um}), (\ref{usb}), (\ref{omb})):
\begin{equation}\label{bsmall}
\sim \frac{6}{\mu^2}(1+\frac{3}{16}g(s)).
\end{equation}

\section*{Conclusion}
The cross-sections and  the slope of the diffraction  are   important global features of hadron interaction dynamics and it is essential that their measurements   can be  directly performed experimentally. 

It is well known that unitarity imposes constraints for the scattering amplitude of the  on mass shell particles 
only. The factorization of the input amplitude  for the fast particles can be interpreted as a manifestation of the independence of transverse and longitudinal dynamics in the first approximation. Their interrelation as a consequence of the unitarization. The  generation of the $B(s)$ energy growth can be treated due to unitarity alone, namely, the unitarization transforms energy independent slope into the one increasing like $\ln^2 s$ at $s\to\infty$. 
This procedure leads, in particular,  to slowing down the asymptotic increase of the total cross--section: instead of violating Froissart--Martin bound by the power--like energy dependence $s^\lambda$, $\lambda>0$, unitarization turns it into a correct $\ln^2 s$ behavior. 

 For the off--shell particles similar unitarity constraints are not applicable without extra assumptions \cite{yndur}. Indeed, there is no Froissart--Martin bound for the case of off--shell particles and unitarity does not rule out an asymptotic power-like behavior of the total cross-sections in this case.

If  unitarity generates the energy dependence of the diffraction cone slope parameter one could expect energy independence of this parameter, for example, in the case of the off--shell particle scattering.
Contrary, the dominance of the Regge mechanism and Pomeron contribution with $\alpha'(0)\neq 0$ assumes ab initio similarity for the on--shell and off-shell scattering processes.

In addition, the structure of the angular distributions of the off-mass shell particle scattering amplitude might be rather different compared to the angular structure of the amplitude for real particles scattering \cite{epjc}.

The studies of the off-mass shell particle scattering in the sophisticated exclusive experiments with three particles in the final state would be definetely helpful for the investigations of the non--perturbative QCD through the soft hadron interactions.

\end{document}